\documentclass[twocolumn,aps,prl,raggedbottom,nobalancelastpage,amssymb,floatfix,longbibliography]{revtex4-1} 
\usepackage[bookmarks=false,linkcolor=blue,urlcolor=blue,colorlinks,citecolor=blue]{hyperref} 
\usepackage{graphicx} 
\usepackage{amsmath} 
\usepackage{amssymb} 
\usepackage{bm} 
\usepackage{pgf} 
\usepackage{color} 
\usepackage{amsfonts} 
\usepackage{hyperref}

\begin{document}

\title{Discriminative Cooperative Networks for Detecting Phase Transitions} 
\author{Ye-Hua Liu} 
\affiliation{Institute for Theoretical Physics, ETH Zurich, 8093 Zurich, Switzerland} 
\affiliation{D\'{e}partement de Physique \& Institut Quantique, Universit\'{e} de Sherbrooke, J1K 2R1 Qu\'{e}bec, Canada} 
\email{yehua.liu@usherbrooke.ca} 
\author{Evert P.L. van Nieuwenburg} 
\affiliation{Institute for Quantum Information and Matter, Caltech, Pasadena, 91125 California, USA} 
\email{evert@caltech.edu}
\begin{abstract}
	The classification of states of matter and their corresponding phase transitions is a special kind of machine-learning task, where physical data allow for the analysis of new algorithms, which have not been considered in the general computer-science setting so far. Here we introduce an unsupervised machine-learning scheme for detecting phase transitions with a pair of discriminative cooperative networks (DCN). In this scheme, a guesser network and a learner network cooperate to detect phase transitions from fully unlabeled data. The new scheme is efficient enough for dealing with phase diagrams in two-dimensional parameter spaces, where we can utilize an active contour model -- the snake -- from computer vision to host the two networks. The snake, with a DCN ``brain'', moves and learns actively in the parameter space, and locates phase boundaries automatically. 
\end{abstract}
\maketitle

The richness of states of matter, together with the power of machine-learning techniques for recognizing and representing patterns, are revealing new methods for studying \emph{emergent phenomena} in condensed matter physics. Paradigms in machine learning have been nicely mapped to those in physics. For example, the classification techniques in machine learning have been applied in detecting classical and quantum phase transitions \cite{Carrasquilla2016,Wang2016,Chng2016,Broecker2016,VanNieuwenburg2017,Schindler2017,Wetzel2017,Wetzel2017a,Ohtsuki2016,Broecker2017,costa17,chng18,Rao:2017ta,Li:2017tu}, the artificial-neural-network architecture has inspired a high-quality Ansatz for many-body wave functions \cite{Carleo2016,Deng2016,Chen2017,Gao2017,Torlai2017,Cai2017,PhysRevB.96.205152,PhysRevX.8.011006}, the generative power of energy-based statistical models is utilized to accelerate Monte Carlo simulations \cite{Wang2017,Huang2016c,Huang2016e,Fujita2017,Liu2017,Liu2016c,Xu2016b,Nagai2017}, and regression has aided material-property prediction \cite{Rupp2012,Rodriguez2013,Pilania2013,Arsenault2014a,Arsenault2015,Bartok2017,Brockherde:2017gz}. Moreover, basic notions from both physics and machine learning can mutually inspire new insights, e.g. a relation between deep learning and the renormalization group \cite{Landon-Cardinal2012,Beny2013,Mehta2014,Lin2016,Rolnick2017,Koch-Janusz2017}.

In physics, the phase (e.g. magnetic vs. non-magnetic phase) is most efficient in summarizing material properties. When changing tuning parameters (e.g. temperature), the material properties may change discontinuously, which is called a phase transition. Machine-learning phase transitions is possible from two angles. In the \emph{supervised} approach, physics knowledge is used to provide answers in limiting cases and the machine learner is asked to extrapolate to the transition point \cite{Carrasquilla2016}. In the \emph{unsupervised} approach, no such knowledge is assumed and the transition is sought by other means \cite{Wang2016,Wetzel2017,costa17,chng18}. 

The confusion scheme proposed previously by us is a \emph{hybrid} method \cite{VanNieuwenburg2017}, where no knowledge of the limiting cases is needed but the learning is still carried out in a supervised manner. Specifically, one first guesses a transition point and tries to train the machine with this guess. When the guess is correct, the machine learner achieves the highest performance. Here we gain the ability to find transitions at the cost of having to repeat the training for many guesses, which is computationally expensive. 

In this work, we extend the confusion scheme by training a ``guesser'' together with the ``learner''. This leads to a fully automated scheme -- the \emph{discriminative cooperative networks} (DCN). In addition, phase transitions in two-dimensional (2D) parameter spaces share many common aspects with image-feature detection in computer vision. However in images the data are the colors, whereas in physics they can be arbitrary results of measurements whose features might not be apparent to the human eyes. This inspires us to use an active contour method \cite{Kass1988}, combined with the DCN scheme, to perform automated searching of phase boundaries in 2D phase diagrams.

We consider data that can be ordered along a tuning parameter $\lambda$. At various values of $\lambda$ the data are described by $\mathbf{d}(\lambda)$, and can be thought of as a vector of real numbers -- results of physical measurements at $\lambda$. We describe a neural network on an abstract level as a map $\mathcal{N}$ that takes data $\mathbf{d}(\lambda)$ and infers the probability distribution $\mathcal{N}(\mathbf{d}(\lambda))=(p_\mathrm{A},p_\mathrm{B},\ldots)$, where $p_i$ represents the probability of $\mathbf{d}(\lambda)$ belonging to phase $i$. Since the data are indexed by $\lambda$, this can be simplified by considering the probability distribution $\mathbf{L}(\lambda)$ directly on $\lambda$. At each $\lambda$ only a single probability (corresponding to the correct phase) should equal to unity, and the rest zero. With phase transitions, the distribution varies with $\lambda$ discontinuously, e.g. for a transition at $\lambda=\lambda_\mathrm{c}$ between two phases A and B there are two components $L_\mathrm{A}(\lambda) = \Theta (\lambda_\mathrm{c}-\lambda)$ and $L_\mathrm{B}(\lambda) = \Theta (\lambda-\lambda_\mathrm{c})$, where $\Theta$ is the Heaviside step-function.

For supervised learning, a large body of $\mathbf{d}(\lambda)$ with the corresponding correct answer $\mathbf{L}(\lambda)$ have to be known beforehand, and the neural network $\mathcal{N}$ is trained with the goal $\mathcal{N}(\mathbf{d}(\lambda))\to\mathbf{L}(\lambda)$. To achieve this goal, parameters $\mathbf{W}_\mathcal{N}$ that characterize the neural network are adjusted during training to minimize a cost function $C[\mathcal{N}(\mathbf{d}(\lambda)), \mathbf{L}(\lambda)]$, quantifying the mismatch between the network's prediction and the known answer. $C$ depends implicitly on the parameters $\mathbf{W}_\mathcal{N}$ through $\mathcal{N}$ and can be minimized using gradient descent methods. 

\begin{figure}
	[t] 
	\begin{center}
		\includegraphics[width=90mm, trim={0.0cm 1.0cm 0.0cm 1.0cm}, clip]{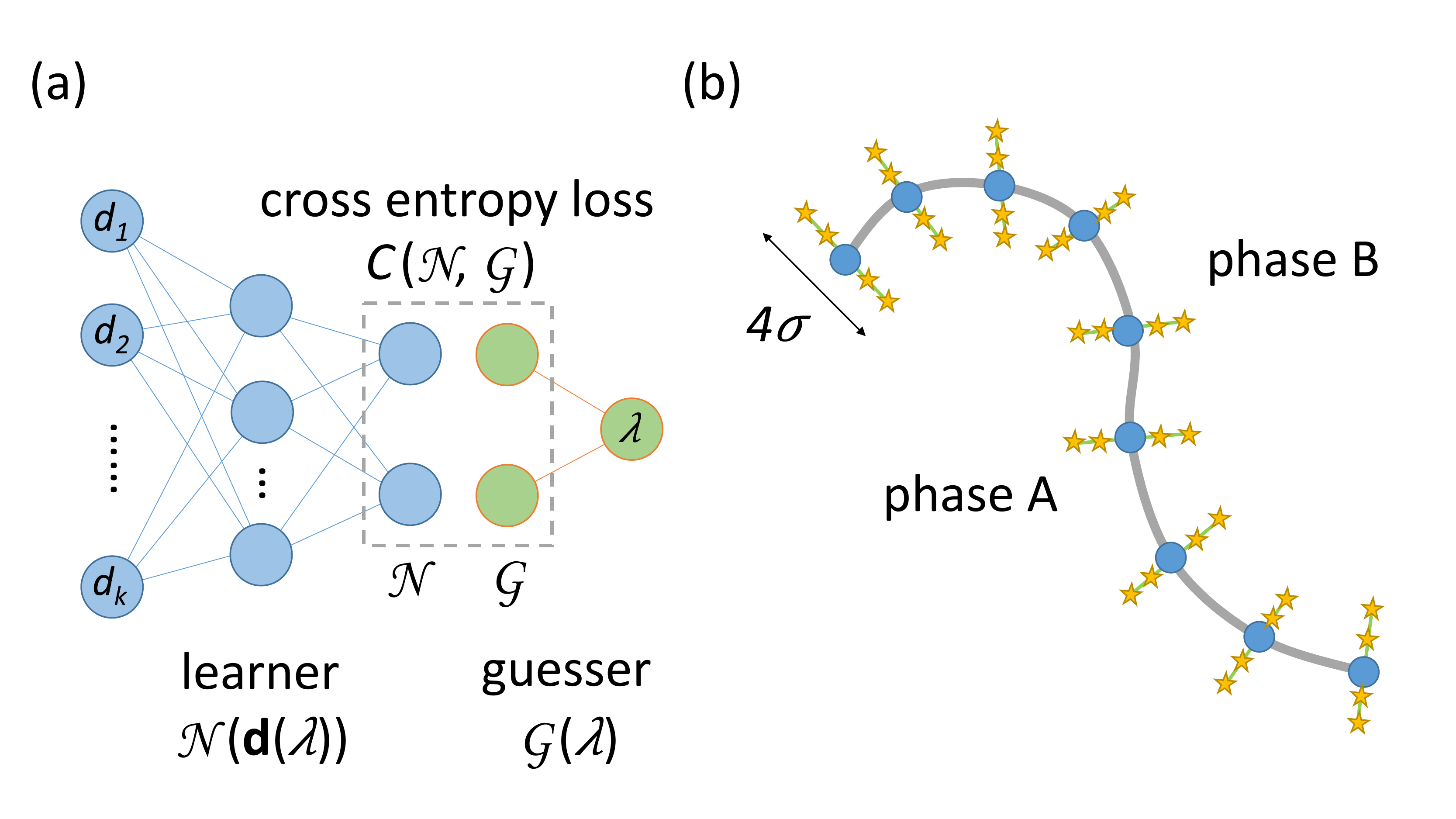} \caption{Schematics for the proposed algorithms. (a) The DCN scheme for learning phase transitions solely from the dataset $\{(\lambda,\mathbf{d}(\lambda))\}$, where $\lambda$ is the tuning parameter and $\mathbf{d}(\lambda)$ is a vector of measurements at $\lambda$. (b) The DCN snake. The blue circles are the snake nodes, the green lines denote normal directions at the nodes. Samples (stars) are generated in the normal direction at each node and are assigned a label according to its distance to the snake. Snakes could be open or closed, and can move to the correct phase boundary (gray line) automatically. The open snake in this figure has 9 nodes and during motion generates mini-batches of training data with mini-batch size $N_b=36$ ($N_b$ is much larger in real simulations). \label{fig:snake}} 
	\end{center}
\end{figure}

Typical machine-learning data live in high-dimensional feature spaces in an \emph{unordered} fashion. The number of ways to separate them into two classes is $2^N$ where $N$ is the size of the dataset. For phase transitions however, all data are \emph{ordered} in the parameter space, and for a single transition point, the number of ways is merely $N+1$. In physics, it is affordable to enumerate all these possibilities to find the most reasonable separation point. This observation led to the confusion scheme \cite{VanNieuwenburg2017}, where one guesses the transition point $\lambda_\mathrm{g}\to\lambda_\mathrm{c}$ and then train the learner network $\mathcal{N}$. By monitoring the number of ``correctly'' classified samples according to this guess -- the performance, the true value for $\lambda_\mathrm{g}$ can be deduced. It turns out the true value is the guess for which the performance is optimal, because here the assigned probabilities in $\mathbf{L}(\lambda)$ and the structures in $\mathbf{d}(\lambda)$ are the most consistent, such that the learner network is \emph{least confused} by the training. 

In the previous proposal, we searched for the optimal $\lambda_\mathrm{g}$ by a brute-force scan of the parameter space. For phase transitions in higher-dimensional parameter spaces, this approach is inefficient. In this work we introduce the guesser network $\mathcal{G}$. It performs the map $\lambda \to \mathcal{G}(\lambda)$, representing the probabilities of $\lambda$ belonging to each possible phase. That is, now the guesser provides $\mathbf{L}(\lambda)$. The guesser is itself characterized by a set of parameters $\mathbf{W}_\mathcal{G}$ on which we wish to perform gradient descent. The overall cost function of the learner $\mathcal{N}$ and guesser $\mathcal{G}$ is now $C\left[\mathcal{N}(\mathbf{d}(\lambda)),\mathcal{G}(\lambda)\right]$, see Fig.~\ref{fig:snake}(a). In this way, we have promoted the human input $\mathbf{L}$ to an active agent $\mathcal{G}$. During training, the learner $\mathcal{N}$ tries to learn the data according to the suggested labels $\mathcal{G}(\lambda)$ obtained from the guesser, and the guesser tries to provide a better set of labels -- they cooperatively optimize the cost $C$.

We first assume one-dimensional (1D) parameter space with two phases, and propose a logistic-regression guesser network with one/two input/output neuron(s): $\mathcal{G}_\mathrm{A,B}(\lambda) = f[s_\mathrm{A,B}(\lambda - \lambda_\mathrm{g})/\sigma]$, where $f(x)=1/(1+e^{-x})$ is the logistic (sigmoid) function, A/B denotes the first/second output neuron, and $s_\mathrm{A,B}=-,+$. The guesser is hence characterized by two parameters $\lambda_\mathrm{g}$ and $\sigma$, setting respectively the guessed transition point and the sharpness of the transition. Gradient descent can be performed on both $\lambda_\mathrm{g}$ and $\sigma$. We use the cross entropy cost function $C(\mathcal{N},\mathcal{G})=-\log\mathcal{N}\cdot\mathcal{G}-\log(1-\mathcal{N})\cdot(1-\mathcal{G})$, which is suitable for classification problems. The gradient of $C$ on the guesser network is obtained by the following equations: 
\begin{eqnarray}
	\frac{ \partial C}{ \partial \mathcal{G}} &=& -\log\mathcal{N} + \log(1-\mathcal{N}), \nonumber\\
	\frac{ \partial\mathcal{G}_\mathrm{A,B}}{ \partial \lambda_\mathrm{g}} &=& -\frac{s_\mathrm{A,B}}{4\sigma\cosh^2\left[(\lambda-\lambda_\mathrm{g})/2\sigma\right]}, \nonumber\\
	\frac{ \partial\mathcal{G}}{ \partial \sigma} &=& \frac{\lambda-\lambda_\mathrm{g}}{\sigma}\frac{ \partial\mathcal{G}}{ \partial \lambda_\mathrm{g}}. 
\label{eq:grads} \end{eqnarray}
These equations fully determine the dynamics of the guesser: $\Delta{\lambda_\mathrm{g}}=-\alpha_{\lambda_\mathrm{g}} \partial C/ \partial {\lambda_\mathrm{g}}$ and $\Delta{\sigma}=-\alpha_\sigma \partial C/ \partial \sigma$, where $\alpha_{\lambda_\mathrm{g}}$ and $\alpha_\sigma$ are the learning rates for the two parameters, respectively. The dynamics of the learner follows $\Delta{\mathbf{W}}_\mathcal{N} = -\alpha_\mathcal{N} \partial C/ \partial \mathbf{W}_\mathcal{N}$ with another independent learning rate $\alpha_\mathcal{N}$, here the gradient is obtained by the back-propagation algorithm \cite{Rumelhart1986}. 

\begin{figure}
	[t] 
	\begin{center}
		\includegraphics[width=85mm]{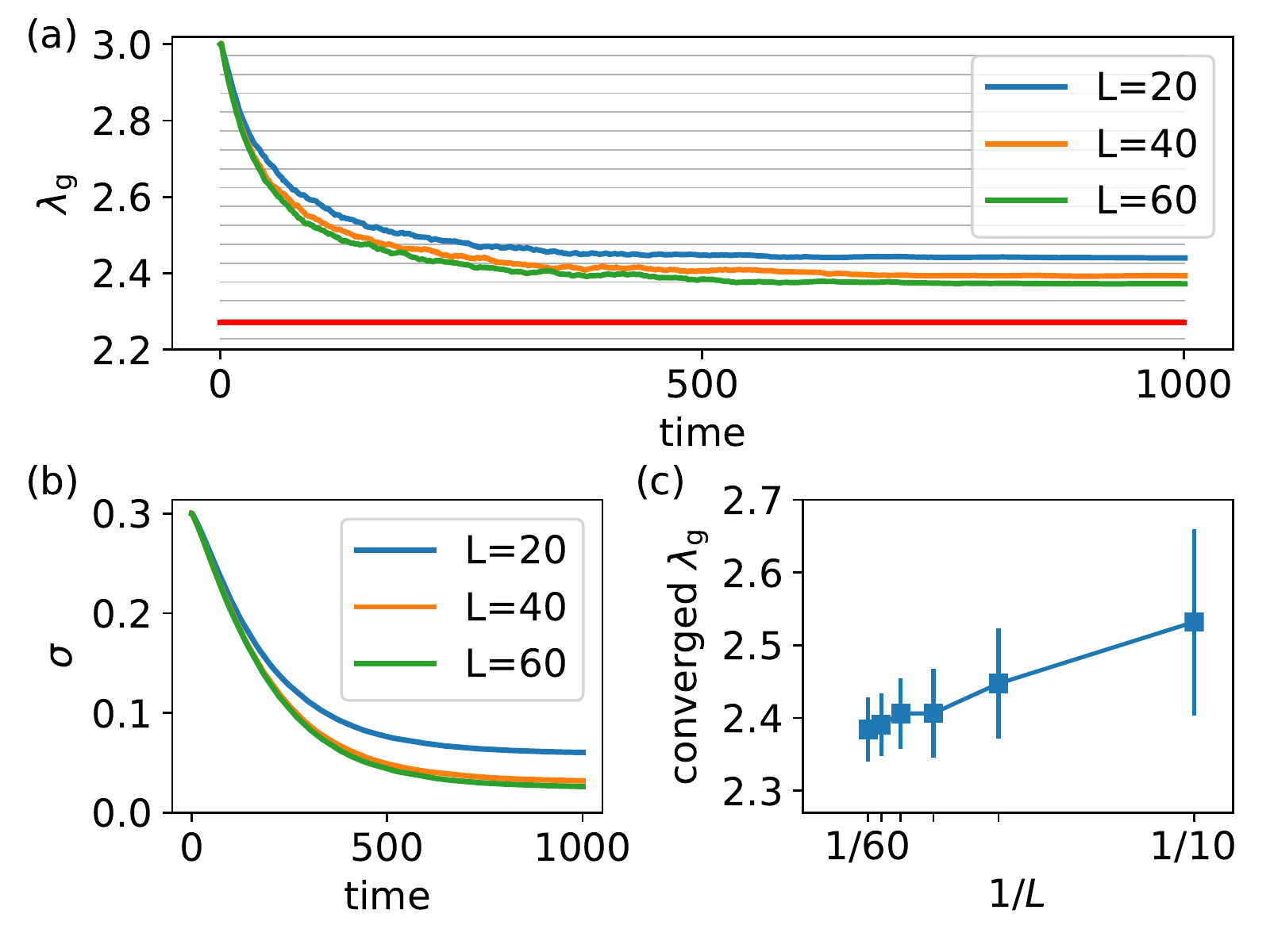} \caption{DCN scheme for the Ising transition. (a) Starting from a higher guess of the transition point, the gradient on the guesser pushes it to move down. The red line marks the exact transition point and the gray lines the temperatures (in the range of the figure) for generating Monte Carlo samples. (b) During training the width $\sigma$ decreases, meaning the combined self-learner is able to distinguish the two phases sharper. Training on samples from larger lattices is faster and more accurate. (c) Finite-size effect on the converged guess $\lambda_\mathrm{g}$, where the length of error bars denotes the converged $\sigma$. Network architecture: fully connected with $L^2$ input neurons, $L$ hidden neurons, and 2 output neurons. Hyper-parameters for training: mini-batch size $N_b=100$; initial learning rates $\alpha_\mathcal{N} = 0.1, \alpha_{\lambda_\mathrm{g}} = 0.025, \alpha_\sigma=0.001$, decay rate $0.995$; dropout keep probability 0.8, $\ell_2$ regularization 0.0001. We have set a lower bound for the width $T>0.01$ and used the mini-batch stochastic gradient descent optimizer \cite{Ruder2016} with batch normalization \cite{ioffe_batch_2015}. \label{fig:ising}} 
	\end{center}
\end{figure}

At this point, one could conceptually regard the guesser and learner together as one compound agent, capable of \emph{self-learning}. We call this scheme \emph{discriminative cooperative networks} (DCN), with the name inspired by the powerful \emph{generative adversarial networks} (GAN) \cite{Goodfellow2014} for generating samples resembling the training data. 

The DCN scheme is efficient because there is no need for repetitive training at each guess. This allows us to move to higher-dimensional parameter spaces. Here we focus on 2D since physics studies usually report phase diagrams in 2D parameter spaces. Inspired by the computer vision techniques for finding image features, we use an active contour model -- the snake \cite{Kass1988} -- for the parametrization of the guesser. 

In computer vision, the snake is a discretized curve of linked nodes, $\mathbf{r}(s)=(x(s),y(s))$, parametrized by $s \in [0,1)$ (for closed snakes) or $s \in [0,1]$ (for open snakes), see Fig.~\ref{fig:snake}(b). The nodes can move actively under ``image forces'', which are the minus gradients of an ``external energy'', with respect to the snake nodes. Specifically, the external energy is the total potential energy $E_\mathrm{external}=\int_0^1\,ds\,\phi(\mathbf{r}(s))$, with the potential $\phi(\mathbf{r})$ proportional to the local color intensity (gradient of color intensity) for line (edge) detection. To keep the snake smooth, internal forces are also introduced, which are derived from the internal energy $E_\mathrm{internal} = \int_0^1\,ds\,\left(\alpha| \frac{\partial \mathbf{r}}{\partial s}|^2+\beta| \frac{\partial^2 \mathbf{r}}{\partial s^2}|^2\right)$. Increasing $\alpha$ makes for a more ``elastic'' snake by preventing stretching and $\beta$ a more ``solid'' snake by preventing bending. The snake evolves in time to lower its total energy $E_\mathrm{total}=E_\mathrm{external}+E_\mathrm{internal}$, and the equation of motion $\dot{\mathbf{v}}\propto-\delta E_\mathrm{total}/\delta \mathbf{r}$ is implemented numerically \cite{Kass1988}. 

In this work, we combine the DCN scheme in artificial intelligence with the snake in computer vision, and the result is an \emph{intelligent snake}. To do this, we replace the conventional image force in computer vision with the machine-learning gradient $\delta E_\mathrm{external}/\delta \mathbf{r}\to \partial C/\partial{\lambda_\mathrm{g}}$. The 1D DCN scheme requires training data from both sides of the guessed transition point. This implies, for the 2D case, a width of the snake. The width, denoted again by $\sigma$, is generically different at each node, and enables the snake to sense its surroundings by selecting training samples in its vicinity within this length scale, as shown in Fig.~\ref{fig:snake}(b). Specifically the sample points are drawn at each node perpendicularly to the snake, with distances uniformly picked in $[-2\sigma,2\sigma]$. The 2D guesser function is then locally the same as in the 1D case, evaluated by each node in its perpendicular direction. For implementation details, see \footnote{The source code can be found in https://github.com/rhinech/snake.}. We note the probing of data within a window (in searching for distinct phases) is a powerful concept that is also successfully used in Ref.~\cite{Broecker2017}.

\begin{figure*}
	[t] 
	\begin{center}
		\includegraphics[height=37mm, trim={1.0cm 0 2.6cm 0}, clip]{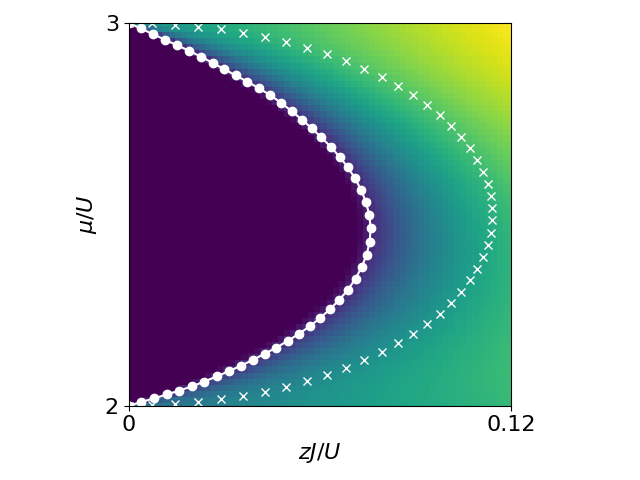} \includegraphics[height=37mm, trim={1.0cm 0 0.0cm 0}, clip]{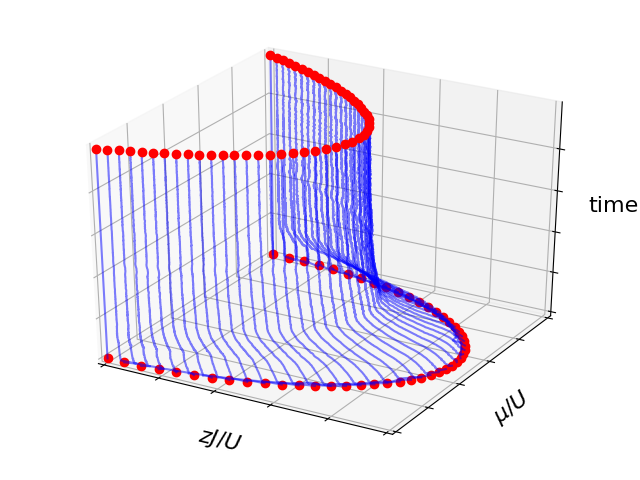} \includegraphics[height=37mm, trim={1.0cm 0 2.6cm 0}, clip]{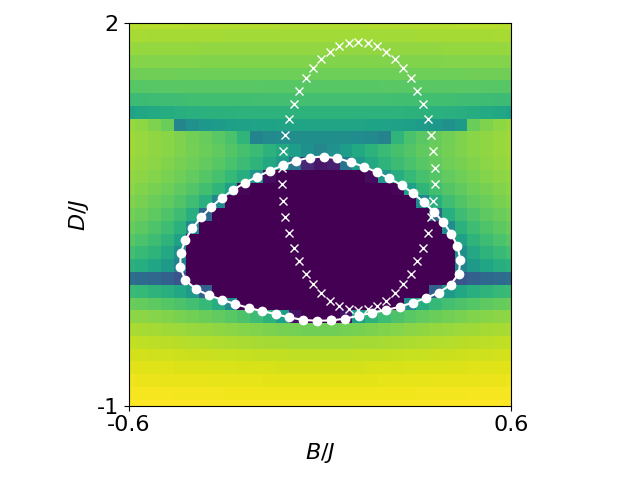} \includegraphics[height=37mm, trim={1.0cm 0 0.0cm 0}, clip]{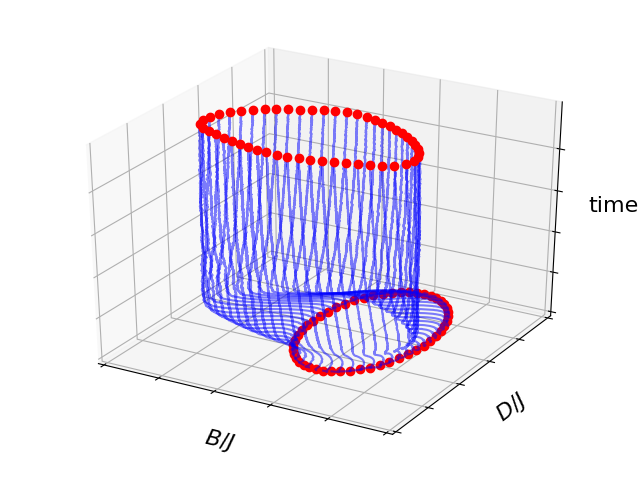} \caption{DCN snake for 2D parameter spaces. (Left panels) The Mott insulator to superfluid transition in the Bose Hubbard model. (Right panels) The topological transition in the spin-1 antiferromagnetic Heisenberg chain. Color plots (purple to yellow goes from zero to nonzero values) show the average hopping for the Bose Hubbard model, and the difference between the largest two eigenvalues of the reduced density matrix for half of the Heisenberg chain. For the Bose Hubbard model we create a large open snake with head and tail fixed at integer chemical potentials $\mu/U=2,3$. In this case the snake shrinks and stops at the correct phase boundary. For the Heisenberg model, we create a large and offset closed snake, which then moves, rotates, shrinks, and finally stays at the Haldane pocket. Parameters for snakes: number of nodes 50; dynamic width at each node is initialized to $T=0.06$ (normalized by the ranges of parameters) and clipped to $0.03<T<0.08$; regularizations $\alpha=0.002$, $\beta=0.4$, $\gamma=0.25$ (see Ref.~\cite{Kass1988} for details). Network architecture: fully connected with 80 input neurons, 80 hidden neurons, and 2 output neurons. Hyper-parameters for training: mini-batch size $N_b=1500$; initial learning rates $\alpha_\mathcal{N}=0.01$, $\alpha_{\lambda_\mathrm{g}}=0.0008$, $\alpha_\sigma=0.0002$, decay rate $0.9999$; dropout keep probability 0.8, $\ell_2$ regularization 0.0001. We used the ADAM optimizer \cite{Kingma2015} because the inputs are sparse for these models \cite{Ruder2016}. \label{fig:hubbard_heisenberg}} 
	\end{center}
\end{figure*}

\paragraph*{Ising model.} We test our scheme on 1D parameter space by studying the classical Ising model on the square lattice: 
\begin{equation}
	H=-J\sum_{\left<i,j\right>}s_i s_j,\quad Z=\sum_{\{s\}}e^{-H/T}, 
\end{equation}
where $s_i=\{-1,1\}$ are the Ising spins, $\left<i,j\right>$ denotes nearest neighbors with coupling $J$, and the tuning parameter is the temperature $T$. This model has a thermal phase transition from the ferromagnetic phase (with aligned spins) to the paramagnetic phase (with random spins) when the temperature is increased across $T_\mathrm{c}\sim2.27J$. The training data $\mathbf{d}(\lambda)=\{s\}_T$ are spin configurations drawn from a Monte Carlo simulation on an $L$ by $L$ square lattice. We select 100 temperatures uniformly from $0.1J$ to $5J$ and prepare 100 samples at each temperature. Every mini-batch consists of $N_b=100$ random samples, one from each temperature \footnote{We have preprocessed each configuration by flipping all its spins when the net magnetization $\sum_i s_i$ is negative.}. Time is measured by the number of learned mini-batches. During training, the guesser moves toward the exact transition point $T_\mathrm{c}\sim2.27J$ and decreases the width $\sigma$ because the discrimination is sharper and sharper (Fig.~\ref{fig:ising}). $\lambda_\mathrm{g}$ does not converge to the exact value when increasing $L$, because the networks most likely learn the order parameter, which is the simplest, but not the sharpest signal for detecting phase transition. It future study we investigate the possibility for the networks to learn also the fluctuations of order parameters.

\paragraph*{Bose-Hubbard model.} As a first example for applying the DCN scheme in 2D parameter spaces, we choose the Bose-Hubbard model: 
\begin{equation}
	H=-J\sum_{\left<i,j\right>}(b^\dagger_i b_j + b^\dagger_j b_i) + \sum_i \left[ \frac{Un_i(n_i-1)}{2} -\mu n_i \right], 
\end{equation}
where $b^\dagger$/$b$ is the bosonic creation/annihilation operator. Regarding the Hubbard interaction $U$ as the energy unit, for each chemical potential $\mu$, the model has a quantum phase transition (at zero temperature) from the Mott insulating state to the superfluid state, when the hopping $J$ is increased \cite{fisher89}. A useful indicator of this phase transition is the average hopping $\left<K\right>$ where $K=\sum_{\left<i,j\right>}(b^\dagger_i b_j+b^\dagger_jb_i)$. Note the notion of $K$ is \emph{unknown to the initial untrained snake}, otherwise the problem reduces to computer vision where machine learning is not needed. The critical point $J_c$ reaches local maxima when the system is at commensurate fillings, corresponding to half-integers $\mu/U$. A phase diagram of this system results in the series of well-known Mott-lobes. We use the mean-field theory developed in Ref.~\cite{Krauth1992} to generate vector data $\mathbf{d}(\lambda_1,\lambda_2)=\mathbf{F}(J,\mu)$, where $F_n$ with $n=0,1,\ldots,\infty$ denotes the amplitude for having $n$ bosons per site, and a cutoff of $n_\mathrm{max}=79$ is chosen for numerics. We target the third Mott lobe with $2\leq\mu/U\leq 3$, and the snake successfully captures the phase boundary as shown in Fig.~\ref{fig:hubbard_heisenberg}. In this case the phase boundary touches the boundary of the parameter space, so we use an open snake with fixed head and tail at known transition points. The snake's motion is then restricted to shrinking or expanding. It is important to emphasize here that we have used knowledge of only two points along the $J=0$ axis in the whole phase diagram, and that the training data seen by the snake is not the average hopping as shown in the background, but the vector data $\mathbf{F}(J,\mu)$ mentioned above \footnote{Additionally, we have tested that the snake is capable of finding the lobe from an initially circular (periodic) configuration.}. 

\paragraph*{Spin-1 Heisenberg chain.} We now move to a quantum phase transition beyond mean-field theory. We choose the spin-1 antiferromagnetic Heisenberg chain with anisotropy and transverse magnetic field: 
\begin{equation}
	H=J\sum_i \mathbf{S}_i \cdot \mathbf{S}_{i+1} + \sum_i \left[D(S^z_i)^2 - B S^x_i\right], 
\end{equation}
where $S_i^a$ are 3 by 3 matrices satisfying $[S_i^a,S_j^b]=i\hbar\delta_{ij}\sum_c\epsilon_{abc}S_i^c$. In the 2D parameter space of magnetic field $B/J$ vs. anisotropy $D/J$, this model has a pocket named the Haldane phase -- a topologically nontrivial phase -- around zero magnetic field and anisotropy \cite{Pollmann2010}. The transition across the boundary of this pocket can be detected by a change in the degeneracy structure of the entanglement spectrum (eigenvalues of the reduced density matrix for part of the spin chain in the ground state), but again the initial untrained snake is unaware of this. For the training data, we simulate an infinite chain with translational invariance using iTEBD \cite{Vidal2007} with bond dimension $m=80$, and record all $m$ eigenvalues $\{\epsilon_1\ldots\epsilon_m\}$ of the reduced density matrix when the chain is cut by half at a bond, i.e. $\mathbf{d}(\lambda_1,\lambda_2)=\{\epsilon\}_{B,D}$. The result is shown in Fig.~\ref{fig:hubbard_heisenberg}. In this model the phase boundary is closed and located near the center of the parameter space. For this reason we use a closed (periodic) snake whose motion now also contains translation and rotation.

In this paper we have proposed the discriminative cooperative networks, capable of self-consistently finding transition points. The high efficiency of this scheme allows us to explore 2D parameter spaces, where we utilized the snake model from computer vision. Our method is in spirit similar to the actor-critic scheme for reinforcement learning \cite{Konda2000} and the adversarial training scheme for generative models \cite{Goodfellow2014}. 

The major limitation for the snake is the need for an initial state that has overlap with the desired features to be detected, so that it is able to probe a gradient. This was also true for their use in computer vision. In applications to phase diagrams, we have the clear advantage of some known extreme limits at which we can fix the snake. We can also overcome this problem by scaling/moving the snake.

Ch'ng et al. have proposed to train neural networks deep inside the known phases with supervision, and then use them to extrapolate the whole phase diagram \cite{Chng2016}. Such a method is, compared to our method, simpler and faster. However the data for supervised training have to be carefully chosen, otherwise interpolation of the phase boundary could be qualitatively incorrect \cite{Chng2016}. On the contrary, the snake can actively explore a much larger area in the parameter space. For general phase transition problems, one could use both methods complementarily.

Machine-learning applications usually assume the existence of big data. However in science it might be expensive to obtain these data. With the DCN scheme, it is possible for a machine-learning agent to suggest parameters for the physicist to carry out experiments/simulations and rapidly locate interesting phenomena. In this paper we put forward a proposal to realize this scheme.

\begin{acknowledgments} The authors thank L. Wang, S. D. Huber, S. Trebst, K. Hepp, M. Sigrist, and T. M. Rice for reading the manuscript and helpful suggestions. Y.-H.L. thank stimulating discussions with G. Sordi and A.-M. Tremblay. Y.-H.L. is supported by ERC Advanced Grant SIMCOFE and the Canada First Research Excellence Fund. E.v.N. gratefully acknowledges financial support from the Swiss National Science Foundation (SNSF) through grant P2EZP2-172185. The authors used TensorFlow \cite{Abadi2016} for machine learning. 
\end{acknowledgments}

\end{document}